\documentclass[11pt]{article}
\usepackage{times}
\usepackage{geometry}
\usepackage[utf8]{inputenc}
\usepackage{enumitem,amssymb}
\usepackage{ragged2e}
\newlist{thematic}{itemize}{8}
\setlist[thematic]{label=$\square$}
\usepackage{pifont}

\usepackage[normalem]{ulem}

\usepackage{fancyhdr}
\usepackage[USenglish]{babel}
\usepackage[utf8]{inputenc}
\usepackage[T1]{fontenc}
\usepackage{epsfig}
\usepackage{comment}
\usepackage{wrapfig}
\usepackage{color}

\usepackage[vflt]{floatflt}
\usepackage{longtable}
\usepackage{caption}
\usepackage{jheppub}   
\usepackage[dvipsnames,svgnames,x11names]{xcolor}
\usepackage[all]{hypcap} 
\usepackage{amssymb,amsmath}
\usepackage{longtable}
\usepackage{amssymb}
\usepackage{amsmath, xspace}
\usepackage{graphics,graphicx} 
\usepackage{rotating}
\usepackage{color}
\usepackage{xcolor}
\usepackage{transparent}
\usepackage{multirow}
\usepackage{subfigure}
\usepackage{sectsty}
\usepackage[outercaption]{sidecap}
\sidecaptionvpos{figure}{c}
\usepackage{titlesec}
\titlespacing*{\section}
{0pt}{1mm}{1mm}
\titlespacing*{\subsection}
{0pt}{1mm}{1mm}

\newcommand{\cc}[1]{{\textcolor{blue}{#1}}} 

\newcommand{\arcsec}{\hbox{$^{\prime\prime}$}}       

\definecolor{DarkGreen}{rgb}{0.0, 0.3, 0.0}
\definecolor{purple}{rgb}{0.5, 0.0, 0.5}
\definecolor{red}{rgb}{1, 0.0, 0.0}
\definecolor{green}{rgb}{0, 1.0, 0.0}

\def\3he{$^3{\rm He}$}

\def\lsim{\mathrel{\lower2.5pt\vbox{\lineskip=0pt\baselineskip=0pt
           \hbox{$<$}\hbox{$\sim$}}}}

\def\gsim{\mathrel{\lower2.5pt\vbox{\lineskip=0pt\baselineskip=0pt
           \hbox{$>$}\hbox{$\sim$}}}}

\usepackage[citestyle=numeric-comp,bibstyle=science,hyperref=true,uniquename=mininit,uniquelist=false,natbib=true,defernumbers=true,eprint=false,sorting=none,maxcitenames=1,maxbibnames=2,]{biblatex}
\makeatletter
\newcommand*{\blx@fnpct@movefor}{}
\newcommand*{\DeclareFootnoteMovePunct}{%
  \@ifstar
    {\let\blx@fnpct@movefor\@empty}
    {}
  \blx@def@fnpct@movefor}
\newcommand*{\blx@def@fnpct@movefor}{%
  \def\do##1{\blx@thecheckpunct{\listadd{\blx@fnpct@movefor}}##1}%
  \docsvlist}
\DeclareFootnoteMovePunct{.,{,}}
\newlength{\blx@fnpct@movelength}
\newcommand*{\blx@fnpct@footnotemover}[1]{%
  #1%
  \ifinlist{#1}{\blx@fnpct@movefor}
    {\settowidth{\blx@fnpct@movelength}{#1}%
     \hspace{-1.\blx@fnpct@movelength}}
    {}%
}
\protected\csedef{blx@acitei@superscript}#1#2#3#4#5{%
  \begingroup
  \blx@citeinit
    \noexpand\blx@fnpct@footnotemover{#5}%
    \blxcitecmd{supercite#1}{#2}{#3}{#4}{}%
  \endgroup}%
\protected\csedef{blx@macitei@superscript}#1#2#3{%
  \noexpand\blx@fnpct@footnotemover{#3}%
  #1{#2}%
  \endgroup}
\makeatother

\DeclareFieldFormat[article]{volume}{#1}
\DeclareFieldFormat[article]{journaltitle}{#1}

\AtEveryBibitem{\clearfield{doi}}
\AtEveryCitekey{\clearfield{doi}}

\AtEveryBibitem{\clearfield{url}}
\AtEveryCitekey{\clearfield{url}}

\AtEveryBibitem{\clearfield{month}}
\AtEveryCitekey{\clearfield{month}}

\DeclareBibliographyDriver{arxiv}{%
    \printnames{author}%
    \newunit\newblock%
    \printfield{eprint}%
    \setunit{\space}%
    \printfield{year}
}

\DeclareFieldFormat[arxiv]{title}{{#1}}
\DeclareFieldFormat[arxiv]{year}{({#1})}
\DeclareFieldFormat[arxiv]{eprint}{arXiv:{#1}}

\renewbibmacro*{name:andothers}{%
  \ifboolexpr
    {
      test {\ifnumequal{\value{listcount}}{\value{liststop}}}
      and
      test \ifmorenames
    }
    {
      \ifnumgreater{\value{liststop}}{1}
        {\finalandcomma}
        {}%
      \andothersdelim
      \bibstring{andothers}%
    }
    {}%
}

\DeclareFieldFormat{labelnumberwidth}{[#1]}

\defbibenvironment{bibliography}
  {}
  {}
  {\addspace
   \printtext[labelnumberwidth]{%
     \printfield{prefixnumber}%
     \printfield{labelnumber}}%
   \addhighpenspace}
   
\renewbibmacro*{finentry}{\finentry\quad}

\begin{document}
\raggedright
\huge
The molecular diversity of the ISM in galaxies across cosmic time
\linebreak
\bigskip
\normalsize


\raggedright
\Large
\normalsize


\bigskip

\textbf{Authors:} 
\textbf{Mathilde Bouvier} (bouvier@strw.leidenuniv.nl, Leiden Observatory, Leiden University, The Netherlands); \textbf{Yiqing Song} (ESO-Chile; MPIfR, Germany); \textbf{Michael Romano} (MPIfR, Germany); Anelise Audibert (IAC, Spain); Ivana Bešlić (LUX, Observatoire de Paris, France); Jakob den Brok (MPIA, Germany); Maria J. Jim\'enez-Donaire (OAN-IGN, Spain; AURA for ESA, STScI, USA); Daizhong Liu (PMO, China), Enrica Bellocchi (UCM, Spain), Matus Rybak (Leiden Observatory, Leiden University, The Netherlands)
\linebreak

\textbf{Science keywords:} galaxies: interstellar medium (ISM); galaxies: evolution; ISM: molecules; astrochemistry \linebreak


 \captionsetup{labelformat=empty}\transparent{0.9}
\begin{figure}[h]
   \centering  
\includegraphics[width=1\textwidth]{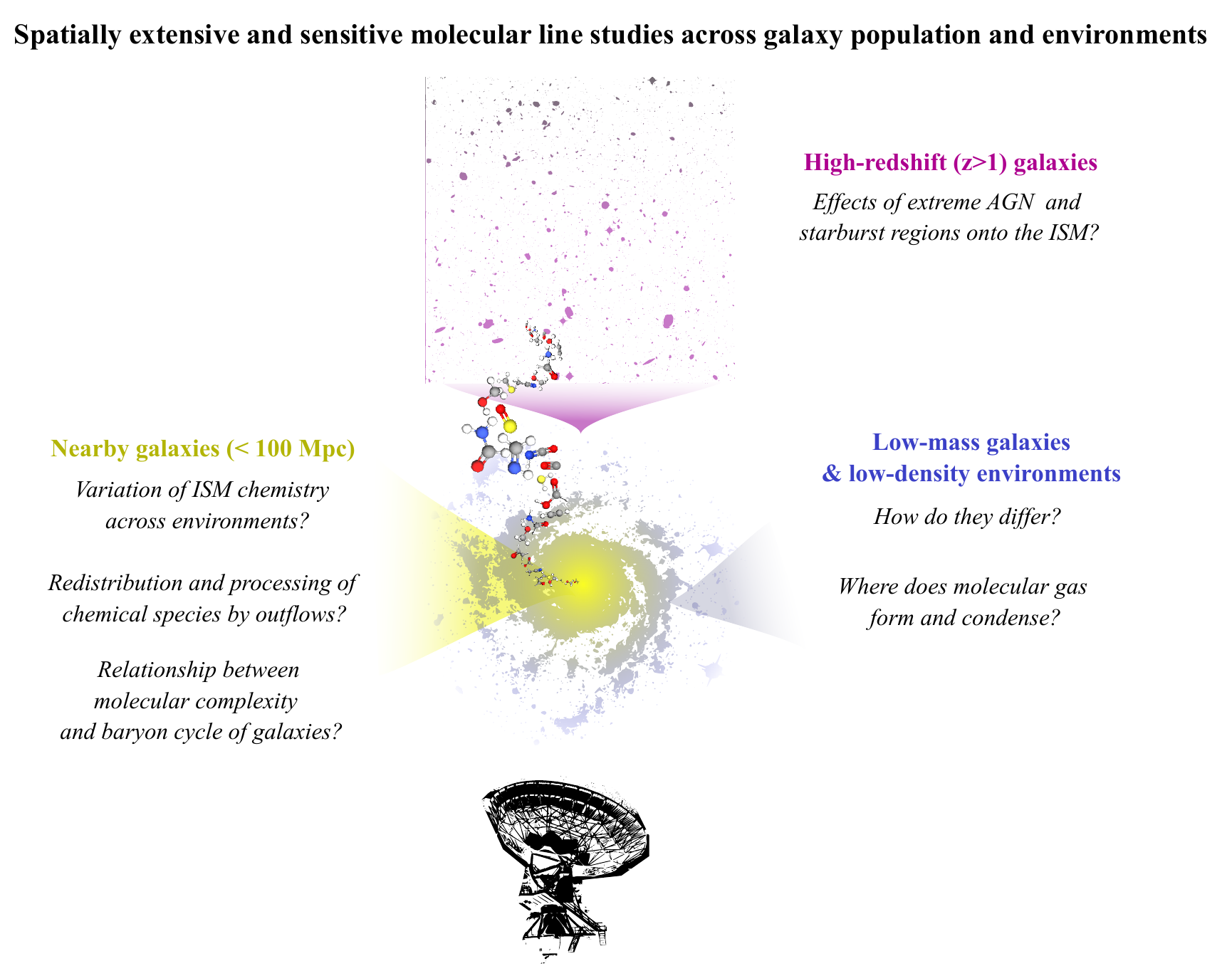}
   \caption{}
\end{figure}
\vspace{-15mm}

\setcounter{figure}{0}
\captionsetup{labelformat=default}


\pagebreak


\transparent{1}
\section*{Abstract}
\begin{justify}
Submillimetre molecular lines (e.g., CO, HCN, SiO) provide a uniquely powerful view of the physical and chemical processes that govern star formation (SF) and galaxy evolution. Yet, our current picture of the molecular universe beyond the Milky Way remains strikingly incomplete: broad chemical inventories exist for only a handful of galaxies, typically more extreme than the Milky Way, constrained by sensitivity limits and narrow survey strategies. In the 2040s, surveying galaxies with multi-species, multi-transitions observations across diverse galactic environments will be crucial to establish effective chemical diagnostics of the various ISM processes from the early universe to $z=0$. 
Extragalactic astrochemistry provides a uniquely sensitive probe of the physical processes shaping galaxies, allowing us to understand, species by species, how gas responds to its local environment and how galaxies grow, transform, and recycle matter over cosmic time.

\section{Scientific context and motivation}\label{sec:intro}


A central challenge in modern astronomy is to understand how galaxies form and evolve. Because galaxy evolution is fundamentally driven by key physical processes unfolding within the interstellar medium (ISM), it is essential to investigate the full molecular reservoir that fuels and regulates these phenomena in different types of systems. Feedback from SF and active galactic nuclei (AGN) injects ultraviolet radiation, X-rays, cosmic rays, and shocks into the surrounding medium, imprinting distinctive physical and chemical signatures onto the molecular gas. These signatures manifest as specific combinations of molecular lines, making molecular spectroscopy one of the most powerful tools for diagnosing ISM conditions (e.g., \citealt{Meier2005, Takano+2014, Viti2017}). \textbf{Deciphering the chemical complexity of the ISM is therefore essential for understanding the diverse processes that shape galaxies across cosmic time.}

Since the first extragalactic detections of CO and OH in the 1970s, advancements in sensitive, broadband ($\geq$10 GHz) (sub-)millimetre (sub-mm) surveys in the early 2000s (e.g., \citealt{Wang2004, Martin+2011, Aladro+2015}) revealed that external galaxies host a molecular richness and complexity comparable to the Milky Way. To date, over 70 molecules have been detected in external galaxies\footnote{\url{https://cdms.astro.uni-koeln.de/classic/molecules}}, proving that the chemical ecosystems of galaxies are far more diverse than once thought. Different molecular species and transitions encode variations in gas density, temperature, metallicity, turbulence, radiation field, and feedback. Thus, tracing their emission across entire galactic disks and outflow regions is an essential diagnostic of the physical mechanisms driving galaxy evolution. Most molecular detections originate from single pointings toward the central regions of only a few nearby galaxies (e.g., \citealt{Wang2004, Nishimura2016, Harada+2018, Takano2019, Sakamoto+2021}) and a handful of high-redshift galaxies (e.g., \citealt{Muller2013, Tercero2020, Yang2023}), limited by line sensitivity, mapping speed, and spectral coverage. Despite this progress, our current view remains incomplete. \textit{To truly understand these relationships, we must move from isolated pointings to wide-field chemical mapping, tracing how molecular composition evolves from galactic centres to outer disks, across dynamically diverse structures, and among different evolutionary and environmental regimes.}



\vspace{-1mm}
\section{Science cases}
Astrochemistry stands at a pivotal moment: the chemical richness of the universe has been revealed, but its structure, diversity, and environmental dependence across galaxy populations remain unknown. Addressing major open questions - \textit{How does the ISM chemistry vary across galactic environments? How galactic outflows redistribute and process chemical species? How do extreme AGN and starburst (SB) regions imprint their energetic signatures onto the ISM? How does the ISM of low-metallicity galaxies and low-density environments differ? What is the relationship between molecular complexity and the baryon cycle of galaxies? Where does molecular gas form and condense?} - requires a new generation of sensitive, broadband, wide-field facilities capable of surveying the molecular universe far beyond the reach of current instruments.

Understanding galaxy evolution requires tracing how molecular gas forms and responds to its local and global environment. While upcoming facilities such as the ELT and SKA will greatly advance our view of stars, ionized gas, and diffuse atomic material, the cold dense phase of the ISM, where the imprint of feedback, energetics, and chemical processing is most directly encoded, remains the least explored. \textbf{The following science cases highlight the key questions that can only be answered through deep, spatially extensive molecular-line studies across different galactic environments and population.}

\subsection{Molecular inventories across cosmic times}

Molecular line surveys at sub-mm wavelengths allow us to observe simultaneously several species and transitions, and provide crucial information about the physical properties of the gas in galaxies. Combining such observations with chemical models, several species were identified as "diagnostic tools" to probe specific environments of the ISM. This is the case for example of CH$_3$OH, HNCO or sulfur-bearing species, which were found to be tracers of large-scale (several tens-hundreds of pc) shocks (e.g., \citealt{saito2017, Huang+2023, bouvier2024}), or CH$_2$NH as a potential probe for compact obscured nuclei\footnote{Compact ($\leq 100$ pc) buried ($ N_{\mathrm{H}_2}\geq 10^{24}$ cm$^{-2}$) nuclei  (e.g., \citealt{Aalto2019}).} (e.g., \citealt{Gorski+2023}). 
Abundance ratios were found to be powerful tools to constrain the cosmic ray ionization rate (HCN/HNC or H$_3$O$^+$/SO; e.g., \citealt{Holdship2022, Behrens+2024}) or to determine the shock history (SiO/HNCO; \citealt{Huang+2023}) in the nearby SB galaxy NGC\,253. Constraining accurately abundances requires access to several transitions of the same species, which is limited by the line sensitivity of the current and planned facilities. To date, there exists only \textit{one} unbiased molecular survey, ALCHEMI (\citealt{Martin+2021}), which mapped the central region of NGC\,253. The survey results in the detection of a wide variety of species and transitions, providing the most complete view of SB environment to date. Other recent line surveys have focused exclusively on a handful near and/or bright galaxies and regions, with limited frequency ranges (e.g., \citealt{Harada+2018, Takano2019, Qiu2020}). Our view of the ISM processes in galaxies is thus still very limited to a few individual sources. \textit{To get significant statistics about the ISM processes and their feedback effects over a broad range of galactic environments, we thus need to expand unbiased molecular line surveys to the broader galaxy population -- a task that will remain severely limited with the current and planned observing capabilities.} 

Studying the ISM in high-redshift ($z\gtrsim1$) galaxies can give us crucial clues about temporal variation of the SF and AGN processes, which were much more intense at $z$=1-4 (e.g., \citealt{Madau2014}). However, typical dense-gas tracers (HCN, HCO$^+$, and HNC) are often ${>}$10$\times$ fainter than CO, making them extremely challenging to detect in high-$z$ galaxies. To date, these tracers have been detected towards only a limited number of high-$z$ galaxies (e.g., \citealt{Walter2003, Wagg2005, Riechers2006, guelin+2007, Bethermin+2018, Canameras+2021, Rybak2022, Rybak2025}).
A handful of high-$z$ galaxies also show molecular richness (e.g., \citealt{Muller2013, Yang2023}), with the detection of even complex species of prebiotic interest (e.g., Formamide, NH$_2$CHO; \citealt{Muller2013}). 
\textit{Expanding dense gas and molecular line surveys towards a larger sample of high-$z$ sources would allow us to fully characterize the high-$z$ dense ISM (hence the most extreme ISM conditions), test physical and chemical theories of the ISM and ultimately, comprehend the chemical evolution and enrichment across cosmic times.} 


\subsection{Probing the full molecular gas budget across galaxy population}

Extensive surveys that probe the dense regions at or in the immediate surroundings of active SF are regularly conducted across a wide range of nearby galaxies (e.g., \citealt{Leroy+2021}). However, these high resolution observations only offer a narrow window into the full baryon cycle, as they generally miss the diffuse cold ISM sensitive to feedback and SF/merger history. To understand how the molecular gas forms and condenses or how it is redistributed via SF- or AGN-feedback driven activity, it is essential to trace also
the diffuse, low-density gas that extends over large galactic scales. 
Sensitive wide-field CO surveys probing down to ${\lesssim} 1\,M_\odot\,$pc$^{-2}$ are required to capture the transition regime between atomic and molecular gas in the galactic outskirts (e.g., \citealt{eibensteiner2024}).
Such observations will also be crucial for unveiling the effects of environmental processes on molecular gas distribution. For example, recent studies using deep single-dish CO(1-0) observations recovering the large scale diffuse emission highlight the necessity of wide-field single-dish mapping surveys of nearby galaxy mergers/clusters to unveil low-density molecular gas shaped by large-scale dynamical processes (e.g., \citealt{Maeda2025}). 
In addition, galactic-scale shocks and turbulence associated with mergers, AGN-driven jets/outflows and SF-driven outflows can effectively heat up the molecular gas and enhance the emission of higher CO transitions (e.g., J$_{\rm up}> 4$; e.g., \citealt{vanderwerf2010,Vallini2019,MontoyaArroyave2024}), which may dominate the global gas excitation of the system 
 (e.g., \citealt{papadopoulos2012}). 
\textit{Sensitive, multi-J CO mapping across a variety of local galaxies are challenging with current and planned facilities, yet are necessary to understand how local physical environments shape molecular gas conditions, and to detect and characterize molecular gas outflows beyond the brightest gas-rich starbursts that have been focused on so far. }  \\
\indent Using CO to trace the bulk of the molecular gas is however not always possible. In particular, in low-mass or low-metallicity galaxies, large fractions of the molecular gas become "CO-dark" as the reduced dust shielding allows far-UV radiation to photodissociate CO. In these environments, it thus becomes important to use alternative tracers such as [CII] (at 1900~GHz) and [CI] (at 492 and 809~GHz) (e.g., \citealt{Liu+2023c, ramambason2024}). 
However, there is still a huge gap towards understanding the roles of [CII] and [CI] in tracing the cold molecular gas and SF feedback and large-scale, spatially-resolved mapping of these tracers in nearby galaxies using current facilities. Ground-based mapping of [CI] at $\lesssim$5\arcsec resolution will be the best way in the next few decades to probe the low-metallicity molecular ISM 
(e.g., \citealt{ramambason2024}) that are prevalent in the early Universe. \textit{By calibrating reliable tracers of the total H$_2$ mass in these environments, we can refine the recipes governing gas-to-star conversion under primordial conditions and thereby improve our understanding of how galaxies form and evolve.} 

\section{Technical requirements}

With the current or planned facilities by 2040, expanding both molecular line surveys towards nearby (D~$<100$~Mpc) and high-$z$ galaxies will remain inaccessible. A new sensitive (sub-)mm facility is required to reach this goal. In particular, we need to be able to \textbf{access the whole (sub-)millimetre wavelength range} ($\sim$30-1000 GHz). This is crucial to target multiple species and transitions and hence derive the gas physical conditions associated with each molecular tracer accurately. Then, we need \textbf{a drastic increase (a factor 10 higher) in the line sensitivity} compared to current or planned facilities to be able to expand unbiased molecular surveys to nearby galaxies, probe the dense ISM in fainter galaxies and environments (e.g., dwarf galaxies or outer regions of galactic disks where the density might be lower) or in high-$z$ galaxies. A \textbf{large ($\sim \mathrm{deg}$) field-of-view}, \textbf{large collecting area} ($\gtrsim$50 m, single-dish), and \textbf{a drastic increase (one order of magnitude) in spectral/spatial mapping speed capabilities} are imperative to survey the wider range of different galaxies and environments, be sensitive to large-scale structures, and recover faint, extended emission in nearby galaxies which can span tens of arcminutes in the sky. \textit{All these technical requirements are necessary for a comprehensive understanding of the dense, multiphase ISM -- hence of the galaxy formation and evolution process -- across the Universe.}

\vspace{-2mm}
{\footnotesize
\printbibliography

@ARTICLE{Liu+2023c,
       author = {{Liu}, Daizhong and {Schinnerer}, Eva and {Saito}, Toshiki and {Rosolowsky}, Erik and {Leroy}, Adam and {Usero}, Antonio and {Sandstrom}, Karin and {Klessen}, Ralf S. and {Glover}, Simon C.~O. and {Ao}, Yiping and {Be{\v{s}}li{\'c}}, Ivana and {Bigiel}, Frank and {Cao}, Yixian and {Chastenet}, J{\'e}r{\'e}my and {Chevance}, M{\'e}lanie and {Dale}, Daniel A. and {Gao}, Yu and {Hughes}, Annie and {Kreckel}, Kathryn and {Kruijssen}, J.~M. Diederik and {Pan}, Hsi-An and {Pety}, J{\'e}r{\^o}me and {Salak}, Dragan and {Santoro}, Francesco and {Schruba}, Andreas and {Sun}, Jiayi and {Teng}, Yu-Hsuan and {Williams}, Thomas},
        title = "{C I and CO in nearby spiral galaxies. I. Line ratio and abundance variations at {\ensuremath{\sim}}200 pc scales}",
      journal = {\aap},
         year = 2023,
        month = apr,
       volume = {672},
          eid = {A36},
        pages = {A36},
          doi = {10.1051/0004-6361/202244564},
archivePrefix = {arXiv},
       eprint = {2212.09661},
 primaryClass = {astro-ph.GA},
       adsurl = {https://ui.adsabs.harvard.edu/abs/2023A&A...672A..36L}
}

@ARTICLE{Leroy+2021,
       author = {{Leroy}, Adam K. and {Schinnerer}, Eva and {Hughes}, Annie and {Rosolowsky}, Erik and {Pety}, J{\'e}r{\^o}me and {Schruba}, Andreas and {Usero}, Antonio and {Blanc}, Guillermo A. and {Chevance}, M{\'e}lanie and {Emsellem}, Eric and {Faesi}, Christopher M. and {Herrera}, Cinthya N. and {Liu}, Daizhong and {Meidt}, Sharon E. and {Querejeta}, Miguel and {Saito}, Toshiki and {Sandstrom}, Karin M. and {Sun}, Jiayi and {Williams}, Thomas G. and {Anand}, Gagandeep S. and {Barnes}, Ashley T. and {Behrens}, Erica A. and {Belfiore}, Francesco and {Benincasa}, Samantha M. and {Be{\v{s}}li{\'c}}, Ivana and {Bigiel}, Frank and {Bolatto}, Alberto D. and {den Brok}, Jakob S. and {Cao}, Yixian and {Chandar}, Rupali and {Chastenet}, J{\'e}r{\'e}my and {Chiang}, I-Da and {Congiu}, Enrico and {Dale}, Daniel A. and {Deger}, Sinan and {Eibensteiner}, Cosima and {Egorov}, Oleg V. and {Garc{\'\i}a-Rodr{\'\i}guez}, Axel and {Glover}, Simon C.~O. and {Grasha}, Kathryn and {Henshaw}, Jonathan D. and {Ho}, I.-Ting and {Kepley}, Amanda A. and {Kim}, Jaeyeon and {Klessen}, Ralf S. and {Kreckel}, Kathryn and {Koch}, Eric W. and {Kruijssen}, J.~M. Diederik and {Larson}, Kirsten L. and {Lee}, Janice C. and {Lopez}, Laura A. and {Machado}, Josh and {Mayker}, Ness and {McElroy}, Rebecca and {Murphy}, Eric J. and {Ostriker}, Eve C. and {Pan}, Hsi-An and {Pessa}, Ismael and {Puschnig}, Johannes and {Razza}, Alessandro and {S{\'a}nchez-Bl{\'a}zquez}, Patricia and {Santoro}, Francesco and {Sardone}, Amy and {Scheuermann}, Fabian and {Sliwa}, Kazimierz and {Sormani}, Mattia C. and {Stuber}, Sophia K. and {Thilker}, David A. and {Turner}, Jordan A. and {Utomo}, Dyas and {Watkins}, Elizabeth J. and {Whitmore}, Bradley},
        title = "{PHANGS-ALMA: Arcsecond CO(2-1) Imaging of Nearby Star-forming Galaxies}",
      journal = {\apjs},
         year = 2021,
        month = dec,
       volume = {257},
       number = {2},
          eid = {43},
        pages = {43},
          doi = {10.3847/1538-4365/ac17f3},
archivePrefix = {arXiv},
       eprint = {2104.07739},
 primaryClass = {astro-ph.GA},
       adsurl = {https://ui.adsabs.harvard.edu/abs/2021ApJS..257...43L}
}

@ARTICLE{Aladro+2015,
       author = {{Aladro}, R. and {Mart{\'\i}n}, S. and {Riquelme}, D. and {Henkel}, C. and {Mauersberger}, R. and {Mart{\'\i}n-Pintado}, J. and {Wei{\ss}}, A. and {Lefevre}, C. and {Kramer}, C. and {Requena-Torres}, M.~A. and {Armijos-Abenda{\~n}o}, R.~J.},
        title = "{Lambda = 3 mm line survey of nearby active galaxies}",
      journal = {\aap},
         year = 2015,
        month = jul,
       volume = {579},
          eid = {A101},
        pages = {A101},
          doi = {10.1051/0004-6361/201424918},
archivePrefix = {arXiv},
       eprint = {1504.03743},
 primaryClass = {astro-ph.GA},
       adsurl = {https://ui.adsabs.harvard.edu/abs/2015A&A...579A.101A}
}

@ARTICLE{Behrens+2024,
       author = {{Behrens}, Erica and {Mangum}, Jeffrey G. and {Viti}, Serena and {Holdship}, Jonathan and {Huang}, Ko-Yun and {Bouvier}, Mathilde and {Butterworth}, Joshua and {Eibensteiner}, Cosima and {Harada}, Nanase and {Mart{\'\i}n}, Sergio and {Sakamoto}, Kazushi and {Muller}, Sebastien and {Tanaka}, Kunihiko and {Colzi}, Laura and {Henkel}, Christian and {Meier}, David S. and {Rivilla}, V{\'\i}ctor M. and {van der Werf}, Paul P.},
        title = "{Neural Network Constraints on the Cosmic-Ray Ionization Rate and Other Physical Conditions in NGC 253 with ALCHEMI Measurements of HCN and HNC}",
      journal = {\apj},
         year = 2024,
        month = dec,
       volume = {977},
       number = {1},
          eid = {38},
        pages = {38},
          doi = {10.3847/1538-4357/ad85db},
archivePrefix = {arXiv},
       eprint = {2409.13821},
 primaryClass = {astro-ph.GA},
       adsurl = {https://ui.adsabs.harvard.edu/abs/2024ApJ...977...38B}
}

@ARTICLE{Bethermin+2018,
       author = {{B{\'e}thermin}, M. and {Greve}, T.~R. and {De Breuck}, C. and {Vieira}, J.~D. and {Aravena}, M. and {Chapman}, S.~C. and {Chen}, Chian-Chou and {Dong}, C. and {Hayward}, C.~C. and {Hezaveh}, Y. and {Marrone}, D.~P. and {Narayanan}, D. and {Phadke}, K.~A. and {Reuter}, C.~A. and {Spilker}, J.~S. and {Stark}, A.~A. and {Strandet}, M.~L. and {Wei{\ss}}, A.},
        title = "{Dense-gas tracers and carbon isotopes in five 2.5 < z < 4 lensed dusty star-forming galaxies from the SPT SMG sample}",
      journal = {\aap},
         year = 2018,
        month = dec,
       volume = {620},
          eid = {A115},
        pages = {A115},
          doi = {10.1051/0004-6361/201833081},
archivePrefix = {arXiv},
       eprint = {1810.04695},
 primaryClass = {astro-ph.GA},
       adsurl = {https://ui.adsabs.harvard.edu/abs/2018A&A...620A.115B}
}

@ARTICLE{Canameras+2021,
       author = {{Ca{\~n}ameras}, R. and {Nesvadba}, N.~P.~H. and {Kneissl}, R. and {K{\"o}nig}, S. and {Yang}, C. and {Beelen}, A. and {Hill}, R. and {Le Floc'h}, E. and {Scott}, D.},
        title = "{Planck's Dusty GEMS. VIII. Dense-gas reservoirs in the most active dusty starbursts at z {\ensuremath{\sim}} 3}",
      journal = {\aap},
         year = 2021,
        month = jan,
       volume = {645},
          eid = {A45},
        pages = {A45},
          doi = {10.1051/0004-6361/202038979},
archivePrefix = {arXiv},
       eprint = {2009.13538},
 primaryClass = {astro-ph.GA},
       adsurl = {https://ui.adsabs.harvard.edu/abs/2021A&A...645A..45C}
}

@ARTICLE{guelin+2007,
       author = {{Gu{\'e}lin}, M. and {Salom{\'e}}, P. and {Neri}, R. and {Garc{\'\i}a-Burillo}, S. and {Graci{\'a}-Carpio}, J. and {Cernicharo}, J. and {Cox}, P. and {Planesas}, P. and {Solomon}, P.~M. and {Tacconi}, L.~J. and {vanden Bout}, P.},
        title = "{Detection of HNC and tentative detection of CN at z = 3.9}",
      journal = {\aap},
         year = 2007,
        month = feb,
       volume = {462},
       number = {3},
        pages = {L45-L48},
          doi = {10.1051/0004-6361:20066555},
archivePrefix = {arXiv},
       eprint = {astro-ph/0612345},
 primaryClass = {astro-ph},
       adsurl = {https://ui.adsabs.harvard.edu/abs/2007A&A...462L..45G}
}

@ARTICLE{Gorski+2023,
       author = {{Gorski}, M.~D. and {Aalto}, S. and {K{\"o}nig}, S. and {Wethers}, C. and {Yang}, C. and {Muller}, S. and {Viti}, S. and {Black}, J.~H. and {Onishi}, K. and {Sato}, M.},
        title = "{The opaque heart of the galaxy IC 860: Analogous protostellar, kinematics, morphology, and chemistry}",
      journal = {\aap},
         year = 2023,
        month = feb,
       volume = {670},
          eid = {A70},
        pages = {A70},
          doi = {10.1051/0004-6361/202245166},
archivePrefix = {arXiv},
       eprint = {2210.04499},
 primaryClass = {astro-ph.GA},
       adsurl = {https://ui.adsabs.harvard.edu/abs/2023A&A...670A..70G}
}

@ARTICLE{Harada+2018,
       author = {{Harada}, Nanase and {Sakamoto}, Kazushi and {Mart{\'\i}n}, Sergio and {Aalto}, Susanne and {Aladro}, Rebeca and {Sliwa}, Kazimierz},
        title = "{ALMA Astrochemical Observations of the Infrared-luminous Merger NGC 3256}",
      journal = {\apj},
         year = 2018,
        month = mar,
       volume = {855},
       number = {1},
          eid = {49},
        pages = {49},
          doi = {10.3847/1538-4357/aaaa70},
archivePrefix = {arXiv},
       eprint = {1801.05941},
 primaryClass = {astro-ph.GA},
       adsurl = {https://ui.adsabs.harvard.edu/abs/2018ApJ...855...49H}
}

@ARTICLE{Huang+2023,
       author = {{Huang}, K.-Y. and {Viti}, S. and {Holdship}, J. and {Mangum}, J.~G. and {Mart{\'\i}n}, S. and {Harada}, N. and {Muller}, S. and {Sakamoto}, K. and {Tanaka}, K. and {Yoshimura}, Y. and {Herrero-Illana}, R. and {Meier}, D.~S. and {Behrens}, E. and {van der Werf}, P.~P. and {Henkel}, C. and {Garc{\'\i}a-Burillo}, S. and {Rivilla}, V.~M. and {Emig}, K.~L. and {Colzi}, L. and {Humire}, P.~K. and {Aladro}, R. and {Bouvier}, M.},
        title = "{Reconstructing the shock history in the CMZ of NGC 253 with ALCHEMI}",
      journal = {\aap},
         year = 2023,
        month = jul,
       volume = {675},
          eid = {A151},
        pages = {A151},
          doi = {10.1051/0004-6361/202245659},
archivePrefix = {arXiv},
       eprint = {2303.12685},
 primaryClass = {astro-ph.GA},
       adsurl = {https://ui.adsabs.harvard.edu/abs/2023A&A...675A.151H}
}

@ARTICLE{Madau2014,
       author = {{Madau}, Piero and {Dickinson}, Mark},
        title = "{Cosmic Star-Formation History}",
      journal = {\araa},
         year = 2014,
        month = aug,
       volume = {52},
        pages = {415-486},
          doi = {10.1146/annurev-astro-081811-125615},
archivePrefix = {arXiv},
       eprint = {1403.0007},
 primaryClass = {astro-ph.CO},
       adsurl = {https://ui.adsabs.harvard.edu/abs/2014ARA&A..52..415M}
}

@ARTICLE{Martin+2011,
       author = {{Mart{\'\i}n}, S. and {Krips}, M. and {Mart{\'\i}n-Pintado}, J. and {Aalto}, S. and {Zhao}, J.-H. and {Peck}, A.~B. and {Petitpas}, G.~R. and {Monje}, R. and {Greve}, T.~R. and {An}, T.},
        title = "{The Submillimeter Array 1.3 mm line survey of Arp 220}",
      journal = {\aap},
         year = 2011,
        month = mar,
       volume = {527},
          eid = {A36},
        pages = {A36},
          doi = {10.1051/0004-6361/201015855},
archivePrefix = {arXiv},
       eprint = {1012.3753},
 primaryClass = {astro-ph.CO},
       adsurl = {https://ui.adsabs.harvard.edu/abs/2011A&A...527A..36M}
}

@ARTICLE{Martin+2021,
       author = {{Mart{\'\i}n}, S. and {Mangum}, J.~G. and {Harada}, N. and {Costagliola}, F. and {Sakamoto}, K. and {Muller}, S. and {Aladro}, R. and {Tanaka}, K. and {Yoshimura}, Y. and {Nakanishi}, K. and {Herrero-Illana}, R. and {M{\"u}hle}, S. and {Aalto}, S. and {Behrens}, E. and {Colzi}, L. and {Emig}, K.~L. and {Fuller}, G.~A. and {Garc{\'\i}a-Burillo}, S. and {Greve}, T.~R. and {Henkel}, C. and {Holdship}, J. and {Humire}, P. and {Hunt}, L. and {Izumi}, T. and {Kohno}, K. and {K{\"o}nig}, S. and {Meier}, D.~S. and {Nakajima}, T. and {Nishimura}, Y. and {Padovani}, M. and {Rivilla}, V.~M. and {Takano}, S. and {van der Werf}, P.~P. and {Viti}, S. and {Yan}, Y.~T.},
        title = "{ALCHEMI, an ALMA Comprehensive High-resolution Extragalactic Molecular Inventory. Survey presentation and first results from the ACA array}",
      journal = {\aap},
         year = 2021,
        month = dec,
       volume = {656},
          eid = {A46},
        pages = {A46},
          doi = {10.1051/0004-6361/202141567},
archivePrefix = {arXiv},
       eprint = {2109.08638},
 primaryClass = {astro-ph.GA},
       adsurl = {https://ui.adsabs.harvard.edu/abs/2021A&A...656A..46M}
}

@ARTICLE{Holdship2022,
       author = {{Holdship}, Jonathan and {Mangum}, Jeffrey G. and {Viti}, Serena and {Behrens}, Erica and {Harada}, Nanase and {Mart{\'\i}n}, Sergio and {Sakamoto}, Kazushi and {Muller}, Sebastien and {Tanaka}, Kunihiko and {Nakanishi}, Kouichiro and {Herrero-Illana}, Rub{\'e}n and {Yoshimura}, Yuki and {Aladro}, Rebeca and {Colzi}, Laura and {Emig}, Kimberly L. and {Henkel}, Christian and {Nishimura}, Yuri and {Rivilla}, V{\'\i}ctor M. and {van der Werf}, Paul P. and {Alma Comprehensive High-Resolution Extragalactic Molecular Inventory (Alchemi) Collaboration}},
        title = "{Energizing Star Formation: The Cosmic-Ray Ionization Rate in NGC 253 Derived from ALCHEMI Measurements of H$_{3}$O$^{+}$ and SO}",
      journal = {\apj},
         year = 2022,
        month = jun,
       volume = {931},
       number = {2},
          eid = {89},
        pages = {89},
          doi = {10.3847/1538-4357/ac6753},
archivePrefix = {arXiv},
       eprint = {2204.03668},
 primaryClass = {astro-ph.GA},
       adsurl = {https://ui.adsabs.harvard.edu/abs/2022ApJ...931...89H}
}

@ARTICLE{Meier2005,
       author = {{Meier}, David S. and {Turner}, Jean L.},
        title = "{Spatially Resolved Chemistry in Nearby Galaxies. I. The Center of IC 342}",
      journal = {\apj},
         year = 2005,
        month = jan,
       volume = {618},
       number = {1},
        pages = {259-280},
          doi = {10.1086/426499},
archivePrefix = {arXiv},
       eprint = {astro-ph/0410039},
 primaryClass = {astro-ph},
       adsurl = {https://ui.adsabs.harvard.edu/abs/2005ApJ...618..259M}
}

@ARTICLE{Muller2013,
       author = {{Muller}, S. and {Beelen}, A. and {Black}, J.~H. and {Curran}, S.~J. and {Horellou}, C. and {Aalto}, S. and {Combes}, F. and {Gu{\'e}lin}, M. and {Henkel}, C.},
        title = "{A precise and accurate determination of the cosmic microwave background temperature at z = 0.89}",
      journal = {\aap},
         year = 2013,
        month = mar,
       volume = {551},
          eid = {A109},
        pages = {A109},
          doi = {10.1051/0004-6361/201220613},
archivePrefix = {arXiv},
       eprint = {1212.5456},
 primaryClass = {astro-ph.CO},
       adsurl = {https://ui.adsabs.harvard.edu/abs/2013A&A...551A.109M}
}

@ARTICLE{Nishimura2016,
       author = {{Nishimura}, Yuri and {Shimonishi}, Takashi and {Watanabe}, Yoshimasa and {Sakai}, Nami and {Aikawa}, Yuri and {Kawamura}, Akiko and {Yamamoto}, Satoshi},
        title = "{Spectral Line Survey toward a Molecular Cloud in IC10}",
      journal = {\apj},
         year = 2016,
        month = oct,
       volume = {829},
       number = {2},
          eid = {94},
        pages = {94},
          doi = {10.3847/0004-637X/829/2/94},
archivePrefix = {arXiv},
       eprint = {1608.01099},
 primaryClass = {astro-ph.GA},
       adsurl = {https://ui.adsabs.harvard.edu/abs/2016ApJ...829...94N}
}

@ARTICLE{Qiu2020,
       author = {{Qiu}, Jianjie and {Zhang}, Jiangshui and {Zhang}, Yong and {Jia}, Lanwei and {Tang}, Xindi},
        title = "{{\ensuremath{\lambda}} = 2 mm spectroscopy observations toward the circumnuclear disk of NGC 1068}",
      journal = {\aap},
         year = 2020,
        month = feb,
       volume = {634},
          eid = {A125},
        pages = {A125},
          doi = {10.1051/0004-6361/201935800},
archivePrefix = {arXiv},
       eprint = {2001.08697},
 primaryClass = {astro-ph.GA},
       adsurl = {https://ui.adsabs.harvard.edu/abs/2020A&A...634A.125Q}
}

@ARTICLE{Riechers2006,
       author = {{Riechers}, Dominik A. and {Walter}, Fabian and {Carilli}, Christopher L. and {Weiss}, Axel and {Bertoldi}, Frank and {Menten}, Karl M. and {Knudsen}, Kirsten K. and {Cox}, Pierre},
        title = "{First Detection of HCO$^{+}$ Emission at High Redshift}",
      journal = {\apjl},
         year = 2006,
        month = jul,
       volume = {645},
       number = {1},
        pages = {L13-L16},
          doi = {10.1086/505908},
archivePrefix = {arXiv},
       eprint = {astro-ph/0605437},
 primaryClass = {astro-ph},
       adsurl = {https://ui.adsabs.harvard.edu/abs/2006ApJ...645L..13R}
}

@ARTICLE{Rybak2022,
       author = {{Rybak}, M. and {Hodge}, J.~A. and {Greve}, T.~R. and {Riechers}, D. and {Lamperti}, I. and {van Marrewijk}, J. and {Walter}, F. and {Wagg}, J. and {van der Werf}, P.~P.},
        title = "{PRUSSIC. I. A JVLA survey of HCN, HCO$^{+}$, and HNC (1-0) emission in z {\ensuremath{\sim}} 3 dusty galaxies: Low dense-gas fractions in high-redshift star-forming galaxies}",
      journal = {\aap},
         year = 2022,
        month = nov,
       volume = {667},
          eid = {A70},
        pages = {A70},
          doi = {10.1051/0004-6361/202243894},
archivePrefix = {arXiv},
       eprint = {2207.06967},
 primaryClass = {astro-ph.GA},
       adsurl = {https://ui.adsabs.harvard.edu/abs/2022A&A...667A..70R}
}

@ARTICLE{Sakamoto+2021,
       author = {{Sakamoto}, Kazushi and {Mart{\'\i}n}, Sergio and {Wilner}, David J. and {Aalto}, Susanne and {Evans}, Aaron S. and {Harada}, Nanase},
        title = "{Deeply Buried Nuclei in the Infrared-luminous Galaxies NGC 4418 and Arp 220. II. Line Forests at {\ensuremath{\lambda}} = 1.4-0.4 mm and Circumnuclear Gas Observed with ALMA}",
      journal = {\apj},
         year = 2021,
        month = dec,
       volume = {923},
       number = {2},
          eid = {240},
        pages = {240},
          doi = {10.3847/1538-4357/ac29bf},
archivePrefix = {arXiv},
       eprint = {2109.08437},
 primaryClass = {astro-ph.GA},
       adsurl = {https://ui.adsabs.harvard.edu/abs/2021ApJ...923..240S}
}

@ARTICLE{Takano+2014,
       author = {{Takano}, Shuro and {Nakajima}, Taku and {Kohno}, Kotaro and {Harada}, Nanase and {Herbst}, Eric and {Tamura}, Yoichi and {Izumi}, Takuma and {Taniguchi}, Akio and {Tosaki}, Tomoka},
        title = "{Distributions of molecules in the circumnuclear disk and surrounding starburst ring in the Seyfert galaxy NGC 1068 observed with ALMA}",
      journal = {\pasj},
         year = 2014,
        month = jul,
       volume = {66},
       number = {4},
          eid = {75},
        pages = {75},
          doi = {10.1093/pasj/psu052},
archivePrefix = {arXiv},
       eprint = {1406.0782},
 primaryClass = {astro-ph.GA},
       adsurl = {https://ui.adsabs.harvard.edu/abs/2014PASJ...66...75T}
}
}

\end{justify}

\end{document}